\documentstyle[aps,epsfig]{revtex}

\newcommand{\ep}{\epsilon}

\newcommand{\pa}{\partial}

\newcommand{\td}{\tilde}
\newcommand{\sla}[1]{\slash\!\!\!\! #1}
\newcommand{\slas}[1]{\slash\!\!\! #1}

\begin{document}

\draft
\title{Noncommutative QED and Muon Anomalous Magnetic Moment}
\author{Xiao-Jun Wang\footnote{E-mail address: wangxj@itp.edu.cn}}
\address{Institution of Theoretical Physics, BeiJing 100080, P.R. China}
\author{Mu-Lin Yan}
\address{Center for Fundamental Physics,
University of Science and Technology of China\\
Hefei, Anhui 230026, P.R. China\footnote{mail
address}}
\date{\today}
\maketitle

\begin{abstract}
The muon anomalous $g$ value, $a_\mu=(g-2)/2$, is calculated up to
one-loop level in noncommutative QED. We argue that relativistic
muon in E821 experiment nearly always stays at the lowest Landau
level. So that spatial coordinates of muon do not commute each
other. Using parameters of E821 experiment, $B=14.5$KG and muon
energy 3.09GeV/c, we obtain the noncommutativity correction to
$a_\mu$ is about $1.57\times 10^{-9}$, which significantly makes
standard model prediction close to experiment.
\end{abstract}

\section{Introduction}
Recently, noncommutative field theory becomes an active subject
again maybe because of development of string/M theory, in which
noncommutative Yang-Mills theory arise in a definite limit of
string theory with a nonzero background field\cite{CDS98,SW99}. In
this sense noncommutativity is motivated by some profound but
beyond the current experiment physics thoughts, such as the
long-held belief that space-time must change its nature at
distances comparable to the Planck scale. The idea of
noncommutativity of spacetime coordinates, however, is quite
old\cite{Snyder47} in physics and mathematics. It also relates to
some today-observable physics. The classic example (though not
always discussed using this language) is the theory of electrons
in a magnetic field projected to the lowest Landau level, which is
naturally thought of as a noncommutative field theory. So far, the
theoretical studies on noncommutative field theory have achieved
a number of results\cite{Douglas01}. It was shown that
noncommutative quantum field theory exhibits an interesting UV-IR
divergence mixing\cite{Mw2000}, and noncommutative gauge theory is
renormalizable and gauge invariant at one-loop level at
least\cite{MS99,MM001}. Furthermore, a noncommutative version of
standard model is builded\cite{MM002}. There are also some
phenomenological studies in framework of noncommutative gauge
theory, such as Aharonov-Bohm effect\cite{MM003}, hydrogen atom
spectrum and the Lamb shift\cite{MM004}.  In this letter, in terms
of these  studies, we will study the muon anomalous magnetic
moment in noncommutative QED.

The anomalous magnetic moments (AMM's) of electron and muon have
been taken as one of the most precise and beautiful tests of the
validity of quantum field theories like QED and the standard
model (SM). They have been calculated and measured to extremely
high precision, and have always been shown a good agreement each
other. This situation, however, seems to has been changed
recently due to E821 experiment at BNL measuring the anomalous
magnetic moment of muon to a precision of 1.3 parts per million
(ppm), which gave a deviation from the SM theoretical value
\begin{eqnarray}\label{1}
\delta a_{\mu}=a_\mu({\rm exp})-a_\mu({\rm SM})=43(16)\times 10^{-10}.
\end{eqnarray}
It is 2.6 times the normal deivation\cite{E821}. This result has
been treated as an indication of new physics and caused extensive
interest in many recent literatures\cite{NP}. Meanwhile, the more
careful theoretical study in framework of SM is still going
on\cite{SMS} for confirming SM prediction. Besides of these
considerations, we should consider the effect of environment of
measure. Sometimes the environment of measure in experiment not
only enters the systematic error, but also changes the physics. A
simplest example is also for an electron motioning in a
(homogeneous, for simple) magnetic field. When electron stays at
the lowest Landau level, position coordinates of electron which
are perpendicular to the magnetic field ${\bf B}$ do not commute
each other (for simple, we assume $B_1=B_2=0,\;B_3=B$=constant)
\footnote{When electron locates at $n^{\rm th}$ Landau level, its
wave function in $x-y$ plane reads $\phi\propto
e^{ik_xx/\hbar}e^{-a^2(y-y_0)^2/2}H_n(a(y-y_0))$ or to
interchange $x$ and $y$. Here $a^2=eB/\hbar c$ and $H_n$ is
Hermite polynomial. Thus $x_0,\ y_0$ denote the center of wave
package only for $n=0$. Meanwhile, precisely the commutative
relation~(\ref{1.1}) should be $[x_0,y_0]=i\frac{\hbar c}{eB}$
instead of arbitrary $x,\ y$. In classical sense, therefore, the
coordinates do not commute only when electron stays at the lowest
Landau level.}
\begin{eqnarray}\label{1.1}
[x^i,x^j]=i\ep^{ij}\frac{\hbar c}{eB},\hspace{0.5in} i,j=1,2.
\end{eqnarray}
It means that electrons in the lowest Landau level should be
described by a non-local field theory rather than usual local
quantum field theory like QED. In usual experiment environment,
electron (or muon) is easily excited to higher Landau level, and
this effect is covered. But when there is large probability that
electron (or muon) stays at the lowest Landau level, the above
effect has to be considered in theoretical prediction.

Now let us focus on the measure of muon AMM in E821 experiment.
There are two main characteristics in E821 experiment: a
homogeneous magnetic field of 14.5KG and highly relativistic muon
with energy 3.09Gev/$c$ ($\gamma_{_{\rm L}}\simeq 29.3$) which
dilates lifetime of muon to $\gamma_{_{\rm L}}\tau\simeq
64.4\mu$s. However, there is effect of synchronous radiation for
circumnutation of highly relativistic muon in magnetic field. It
forces that muon loses energy ceaselessly and nearly always stays
at the lowest Landau level. In this sense, muon physics in this
situation should be described by a non-local quantum field theory
(in particular, noncommutative QED) instead of usual QED. In this
letter, therefore, noncommutative QED will be used to calculate
muon AMM in E821 experiment.

The letter is organized as follows. In sect. 2, the basic notation
of noncommutative field theory is reviewed. In sect. 3, we
calculate the muon AMM in noncommutative QED to one-loop, and
correction on muon AMM due to noncommutative coordinates is
obtained. In sect. 4, we first evaluate the numeric result of the
noncommutative correction. Then we devote a brief summary.

\section{Noncommutative Quantum Electrodynamics in ${\cal M}^4$}

Consider four dimension space-time with coordinates
$x^{\mu},\;\mu=0,...,3$ which obey the following commutation
relations
\begin{eqnarray}\label{2.1}
[x^\mu,x^\nu]=i\theta^{\mu\nu},
\end{eqnarray}
where $\theta^{\mu\nu}$ is a constant asymmetric tensor. In
particular, in this letter we focus on only spatial coordinates
are noncommutative, i.e., $\theta^{0i}=0,\;\theta^{ij}\neq
0,\;i,j=1,2,3$. By noncommutative space-time one means the
algebra ${\cal A}_\theta$ generated by the $x^\mu$ satisfying
(\ref{2.1}), together with some extra conditions on the allows
expressions of the $x^\mu$. The elements of ${\cal A}_\theta$ can
be identified with ordinary functions on ${\cal M}^4$, with the
product of two functions $f$ and $g$ given by the Moyal formula
(or star product):
\begin{eqnarray}\label{2.2}
(f\star g)(x)={\rm exp}[\frac{i}{2}\theta^{\mu\nu}
  \frac{\pa}{\pa x_1^\mu}\frac{\pa}{\pa x_2^\nu}]f(x_1)
  g(x_2)|_{x_1=x_2=x}
\end{eqnarray}
A field theory is defined as usual by constructing an action, but
replace ordinary product by star product. For example, the action
of noncommutative quantum electrodynamics (NCQED) is
\begin{eqnarray}\label{2.3}
S=\int d^4x\{\bar{\psi}\star(i\sla{\pa}-m)\star\psi+e\bar{\psi}\star\sla{A}\star\psi
 -\frac{1}{4}F^{\mu\nu}\star F_{\mu\nu}\}.
\end{eqnarray}
This action exhibits an U(1) gauge symmetry, and the gauge field
is provided by a real vector function, $A_\mu(x)$, on ${\cal
M}^4$. But field strength $F_{\mu\nu}$ for this gauge field now
reads $F_{\mu\nu}=\pa_\mu A_\nu-\pa_\nu
A_\mu-ie[A_\mu,A_\nu]_{\star}$, where
$[A_\mu,A_\nu]_{\star}=(A_\mu\star A_\nu)(x)-(A_\nu\star
A_\mu)(x)$. Thus U(1) noncommutative QED is similar to usual
Yang-Mills theory.

It has been shown that the basic structure of renomalization of
noncummutative field theory is rather different from usual
commutative gauge theory. In other words, the UV properties are
controlled by the planar diagrams, while nonplanar diagrams
generally lead through what is called "UV/IR mixing"\cite{Mw2000}
to new IR phenomena. The limit $\theta\rightarrow 0$ in these
theories is non-analytic. Meanwhile, pure U(N) noncommutative
gauge theory is  renormalizable and gauge invariant, at least at
one-loop level: U(1) case is studied in ref.\cite{MS99,MM005}, and
general U(N) case is studied in
refs.\cite{Armoni2001,KT2001,Zanon2001a,Ruiz2001}. In addition,
it has been also shown that quantum noncommutative field theory
is unitary theory if it defines in Euclidean space or
noncommutatiivity is purely spatial ($\theta_{0i}=0$) in
Minkowski space-time. Therefore, noncommutative QED considering
by this letter is well-defined even at quantum level.

The Feynman rules for NCQED (Feynman-'t Hooft gauge)
reads\cite{MM005}

\vspace{-0.1in}
\parbox{3in}{\mbox{}\\ \mbox{} \\ \mbox{} \\
\begin{figure}[hp]
  \hspace{1in}\psfig{figure=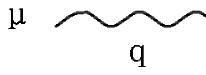,width=0.7in}

  \hspace{1.1in}\psfig{figure=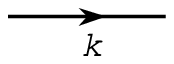,width=1.2in}

  \hspace{1.2in}\psfig{figure=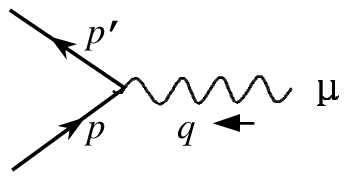,width=1.2in}

  \hspace{1.2in}\psfig{figure=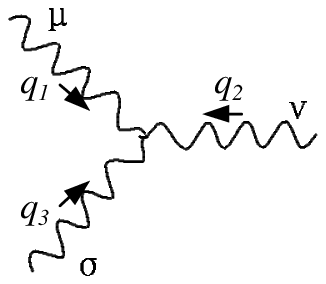,width=1.2in}
\end{figure}}
\hfill
{\parbox{3in}{\begin{eqnarray*} &&\frac{-ig_{\mu\nu}}{q^2+i\ep} \\
&&\mbox{}\\ &&\frac{i}{\slas{k}-m+i\ep} \\ &&\mbox{}\\ &&\mbox{}\\
&&ie\gamma^\mu e^{-\frac{i}{2}p\times_{\theta}p'} \\ &&\mbox{}\\
&&\mbox{}\\ &&ie(e^{-\frac{i}{2}q_2\times_{\theta}q_3}-
e^{\frac{i}{2}q_2\times_{\theta}q_3})\{g_{\mu\sigma}(q_1-q_3)_\nu
\\ &&+g_{\mu\nu}(q_2-q_1)_\sigma+g_{\nu\sigma}(q_3-q_2)_\mu\}
\end{eqnarray*}}
\mbox{}\\
where $p\times_{\theta}k=p_\mu\theta^{\mu\nu}k_\nu$. Feynman rules
for other vertices, such as four-photon vertex and ghost vertex,
are independent of AMM of fermion. So that we ignore them in this
letter.

\section{Fermion vertex function in NCQED}

Up to one-loop level, there are two one-particle irreducible
diagrams which contribute to fermion vertex function (fig. 1-(a)
and fig. 1-(b)).
\begin{figure}[hpbt]
  \centering{\psfig{figure=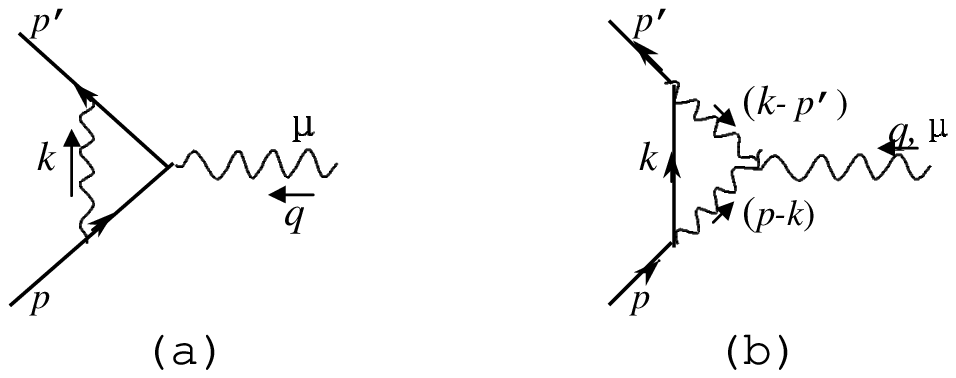,width=3.5in}}
\begin{minipage}{5in}
\caption{One-loop correction to fermion vertex function in NCQED}
\end{minipage}
\end{figure}}

Direct calculation shows that contribution from fig. 1-(a) is same
to usual commutative QED, i.e., there is no nonplanar diagrams in
fig. 1-(a). However, the fig. 2-(a) entirely new. It denotes
contribution from noncommutativity at one-loop level, and corrects
fermion vertex function of usual QED. Since there is additional
tensor $\theta^{\mu\nu}$, general Lorentz structure of vertex
function is rather complicated in NCQED. Concretely, the formal
structure of fermion vertex function $\delta\Gamma^\mu$, which is yielded
by fig. 1-(b), is follow
\begin{eqnarray}\label{3.1}
e^{\frac{i}{2}p\times_{\theta}p'}\delta\Gamma^\mu &=&
  H_1\gamma^\mu+\frac{i}{2m}H_2\sigma^{\mu\nu}q_\nu
  +\frac{i}{2m}H_3q^\mu+imH_4\td{q}^\mu\nonumber \\
&&+H_5(p+p')^\mu\td{q}\cdot\gamma
    +H_6\td{q}^\mu\td{q}\cdot\gamma,
\end{eqnarray}
where $\td{q}^\mu=q_\sigma\theta^{\sigma\mu}$, and form factors
$H_i(i=1,...,6)$ are function of Lorentz scalars $q^2,\;q\circ
q\equiv
g_{\rho\sigma}q_\mu\theta^{\mu\rho}\theta^{\sigma\nu}q_\nu$~
\footnote{Our definition on $q\circ q$ has an opposite sign
comparing with some references, since here we use metric
$(+,\;-,\;-,\;-)$.} and $p\times_\theta p'$ . Let us see which
form factors will contribute to fermion magnetic moment (we can
conveniently take fermion rest framework, such that
$p\times_\theta p'=0$ for $\theta^{0i}=0$):
\begin{enumerate}
\item Conservation of electromagnetic current requires
$H_3(q^2=0,q\circ q)=0$. Direct calculation will also shows this
point.
\item The form factor $H_1(q^2,q\circ q)$ relates to normalization of
electronic charge. In other word, if general vertex function is
$$e^{\frac{i}{2}p\times_{\theta}p'}\delta\Gamma^\mu=
  F_1\gamma^\mu+{\rm other\;independent\;terms},$$
normalization of electronic charge requires $F_1(q^2=0,q\circ
q)=1$, where $F_1(q^2,q\circ q)$ sum all Feynman diagram
contribution and $H_1$ is included in $F_1$.
\item Due to $q_\sigma\theta^{\sigma\mu}A_\mu\rightarrow
\vec{\theta}\cdot\vec{B}$ for $\theta^{0i}=0$, the $H_4$ term is
independent of fermion spin in non-relativistic limit since the
bilinear structure $\bar{u}(p')u(p)$. This term denotes a
magnetic mass from effect of noncommutativity \footnote{The many
expression of this section can be recasted from one in
ref.\cite{MM005}. However, it distinguishes from our discussion
that authors of ref.\cite{MM005} thought the $H_4$ term should
contribute to muon AMM. This difference may be distinguished in
future high precision experiment on fermion mass.}.
\item The $H_6$ term can be removed by means of field
redefinition
\begin{eqnarray}\label{3.2}
A_\mu\longrightarrow A_\mu'=A_\mu+\frac{H_6}{2F_1}
\theta^{\rho\sigma}\theta_{\mu\alpha}D^{\alpha}F_{\rho\sigma}.
\end{eqnarray}
\item Using Dirac equation for external line fermion, we
can obtain
\begin{eqnarray}\label{3.3}
\frac{1}{2}(p'-p)_\nu\td{q}_\rho\bar{u}(p')[\gamma^\mu,\gamma^\nu]
 \gamma^\rho u(p)=(p+p')^\mu\bar{u}(p')\td{q}
 \cdot\gamma u(p)+4(p\times_\theta p')\bar{u}(p')\gamma^\mu u(p).
\end{eqnarray}
Then we have
\begin{eqnarray}\label{3.4}
(p+p')^\mu\bar{u}(p')\td{q}\cdot\gamma u(p)\Longrightarrow
  i\ep^{\mu\nu\sigma\rho}q_\nu \td{q}_\sigma\bar{u}(p')
  \gamma_\rho\gamma_5 u(p).
\end{eqnarray}
It is easy to check that $\bar{u}(p')\gamma_i\gamma_5
u(p)\;(i=1,2,3)$ is independent of fermion spin operator at
non-relativistic limit, i.e., the indices $\mu,\;\nu,\;\sigma$ in
right side of eq.~(\ref{3.4}) should be spatial indices. Thus for
very slowly varying external vector potential, $A_\mu^{\rm
cl}(x)=(0,\vec{\bf{A}}^{\rm cl}(\vec{\bf{x}}))$, we have
\begin{eqnarray}\label{3.5}
(p+p')^\mu A_\mu\bar{u}(p')\td{q}\cdot\gamma
u(p)\hspace{0.1in}\propto\hspace{0.1in}
\ep^{jl}\ep_{ijk}q_iq_lA_k\hspace{0.1in}\propto\hspace{0.1in}
\sum_k (\pa^2A_k-\pa_i\pa_k A^i)=0,
\end{eqnarray}
due to equation of motion of external electromagnetic field.
Therefore, the $H_5$ term also does not contribute to fermion
magnetic moment.
\end{enumerate}
According to the above analysis, we can conclude that
noncommutativity correction to fermion AMM is
\begin{eqnarray}\label{3.6}
\delta a_f=H_2(q^2=0,q\circ q).
\end{eqnarray}

Applying the Feynman rules in sect. 2, we find, to order $\alpha$
and for small $m^2q\circ q$,
\begin{eqnarray}\label{3.7}
H_1(q^2=0,q\circ q)&=&-\frac{3\alpha}{8\pi}[\frac{2}{\ep}
   +\ln{(\mu^2q\circ q)}]+{\cal O}(1), \nonumber \\
H_2(q^2=0,q\circ q)&=&-\frac{\alpha}{96\pi}m^2q\circ q
  [\ln{\frac{m^2q\circ q}{4}}+2\gamma_{_{\rm E}}-\frac{19}{6}]
  +{\cal O}((m^2q\circ q)^2),\\
H_4(q^2=0,q\circ q)&=&-\frac{\alpha}{24\pi}
  [\ln{\frac{m^2q\circ q}{4}}+2\gamma_{_{\rm E}}-\frac{8}{3}]
  +{\cal O}(m^2q\circ q),\nonumber
\end{eqnarray}
where $\mu$ is scale factor of dimensional regularization,
$\gamma_{_{\rm E}}\simeq 0.5772$ is Euler constant. The full
expression on these form factors can be found in
ref.~\cite{MM005}. The UV divergence for $\ep\rightarrow 0$ in
$H_1$ is from planar diagram, and IR divergence for $q\circ
q\rightarrow 0$ (precisely it should be called "UV/IR mixing") in
$H_1$ and $H_4$ are from nonplanar diagram. Fortunately, there is
neither UV nor IR divergence in $H_2$. Thus noncommutativity
correction to fermion AMM is finite at one-loop level. In
addition, direct evaluations show that there are no further UV
divergence in $H_5$ and $H_6$, thus NCQED are also renormalizable
up to one-loop.

\section{Noncommutativity correction to muon AMM}

Now let us return to muon AMM in E821 experiment. It is convenient
to take the homogeneous magnetic field is along $x^3$ direction.
Then noncommutative parameters $\theta^{\mu\nu}$ are given by
\begin{eqnarray}\label{4.1}
\theta^{0i}=0,\hspace{0.5in}\theta^{13}=\theta^{23}=0,\hspace{0.5in}
\theta^{12}=-\theta^{21}=\theta=\frac{\hbar c}{eB}.
\end{eqnarray}
In particular, in muon rest framework the parameter $\theta$ is
dilated to $\theta=\hbar c/(eB\gamma_{_{\rm L}}^2)\simeq 5.3\times
10^{-15}{\rm cm}^2\simeq 1.36\times 10^7{\rm MeV}^{-2}$, where
magnetic induction $B\simeq 14.5K$G, and $\gamma_{_{\rm L}}\simeq
29.3$ is Lorentz factor.

In microscopic description of interaction between external
magnetic field and muon, the dominant effect is that very low
energy photon is scattered by relativistic muon (inverse Compton
scattering), and photon obtain higher energy after scattering
(synchronous radiation in classical electrodynamics). Since muon
always loses energy in this mechanism, it stays at the lowest
Landau level in the most time. In addition, the inverse Compton
scattering tell us $q_3\simeq 0$ in muon rest framework (where
$q_\mu=(E_\gamma,\vec{q})$ denotes four-moment of incident
photon). Then we have $q\circ q\simeq \theta^2E_\gamma^2$. It is
impossible to exactly evaluate numeric result of noncommutativity
correcion to muon AMM. The reason is that we can not know
$E_\gamma$ exactly, or rather, $E_\gamma$ distributes in large
region. In this sense, the noncommutativity correction to muon AMM
is statistical. From inverse Compton scattering we have
$E_\gamma\sim E^{^{SR}}_\gamma/\gamma_{_{\rm L}}^2$, where
$E^{^{SR}}_\gamma$ is photon energy in synchronous radiation). In
addition, the spectrum distribution function of synchronous
radiation in $x^1-x^2$ plane is well-known
\begin{eqnarray}\label{4.2}
N_{\Delta\omega}(\omega^{^{SR}})\;\propto\;\omega^{^{\rm SR}}
K_{2/3}^2(y/2),
\hspace{1in} y=\frac{\omega^{^{\rm SR}}}{\omega_c},
\end{eqnarray}
where $K_{2/3}$ is the modified Bessel function of the second kind,
$\omega_c$ is critical frequency of synchronous radiation,
\begin{eqnarray}\label{4.3}
\omega_c=\frac{3eB}{2m_\mu c}\gamma_{_{\rm L}}^2\Longrightarrow
(E^{^{SR}}_\gamma)_c=\frac{3\hbar eB}{2m_\mu c}\gamma_{_{\rm L}}^2.
\end{eqnarray}
The above equation yields that critical energy of incident photon
is $E_\gamma^c\sim 3\hbar eB/(2m_\mu c)\simeq 1.2\times
10^{-12}$MeV, which is indeed very small. Then using
eq.~(\ref{3.7}), we obtain the noncommutativity correction to muon
AMM as follow
\begin{eqnarray}\label{4.4}
\delta a_\mu&\simeq&-\frac{\alpha}{96\pi}\frac{m_\mu^2\theta^2
(E^{^{SR}}_\gamma)_c^2}{\gamma_{_{\rm L}}^4}
\frac{\int_0^\infty y^3\{\ln{\frac{m_\mu^2\theta^2
(E^{^{SR}}_\gamma)_c^2}{4\gamma_{_{\rm L}}^4}}+2\gamma_{_{\rm E}}
-\frac{19}{6}+\ln{y^2}\}K_{2/3}^2(y/2)dy}
{\int_0^\infty yK_{2/3}^2(y/2)dy} \nonumber \\
&=&-\frac{9\alpha}{384\pi\gamma_{_{\rm L}}^4}
\frac{\int_0^\infty y^3\{\ln{\frac{9}{16\gamma_{_{\rm L}}^4}}
+2\gamma_{_{\rm E}}-\frac{19}{6}+\ln{y^2}\}K_{2/3}^2(y/2)dy}
{\int_0^\infty yK_{2/3}^2(y/2)dy} \nonumber \\ &=&1.57\times
10^{-9}.
\end{eqnarray}
It is surprise that noncommutativity correction makes SM
prediction close to experiment, and also is order to $10^{-9}$.

We shall conclude with several remarks. First, our study in this
letter is heuristic rather than an ultimate conclusion. The
quantum mechanics states that, when a charge particle is in the
lowest landau level, coordinates of center of its wave package
(not any spatial coordinates) do not commute. Meanwhile, the
noncommutative field theory used by this letter has a prior
assumption that arbitrary coordinates in different directions
fail to commute. Therefore, the noncommutative field theory only
is an approximate description on single charge particle lying the
lowest Landau level. We still do not know how to evaluate error
bar of the approximation. Of course, the higher energy of
particle (or the shorter its Compton wave length) is, the better
this approximation is.

Secondly, it is well-known that noncommutativity also originates
from decoupling limit of string theory, or it is an intrinsic
property of spacetime rather than induction of proper background
field. This effect, however, will correct muon AMM in very tiny
order of magnitude. For example, if we assume that length of
noncommutativity is smaller than classical electron radius, i.e.,
$\theta<10^{-30}$cm$^2=10^{-8}$MeV$^{-2}$, the correction to muon
AMM will be smaller than $10^{-37}$.  Indeed, the value of
$\theta$ obtained from Lamb shift\cite{MM004} is
$\theta<10^{-8}$GeV$^{-2}$, which is much smaller than our
evaluation. This large difference can be easily interpreted: In
measure of Lamb shift, electron in Hydrogen atom is no longer
relativistic. It is easily excited to higher Landau level. So
that if noncommutativity contributes to Hydrogen atom spectrum
and Lamb shift, it must have other origination (string theory? or
intrinsic property of spacetime?) instead of the lowest landau
level considered by this present letter.

Thirdly, from the second line of eq.~(\ref{4.4}), we can see that
noncommutativity correction to fermion AMM is independent of mass
of fermion in relativistic limit. It implies that, if
configuration proposed by this letter is right, to measure
electron AMM in environment of E821 experiment, we will find same
correction on electron AMM. This can also test whether
noncommutative field theory is a good approximation to describe
single fermion lying the lowest Landau level.

There are three and four photon vertices in NCQED. However, we should
remember that, in our consideration the noncommutativity
locates in coordinates of position of fermion. So that three and
four photon vertices are virtual, and only exist in fermionic
interaction. It is impossible that all photon are on-shell in these
vertices.

Finally we note that there are some theoretical problems on NCQED
needed to be solved: direct renormalization calculation and $\beta$
function of NCQED; infrared safety for limit $q\circ q\rightarrow
0$, etc. These problems will be studied in forthcoming papers.

The authors acknowledge professor M.M. Sheikh-Jabbari for
providing us useful literature on NCQED which is very important
for our studies. They also thank professors Y.-L. Wu and M. Li for
discussions which clarifying some physical understanding of our
results. This work is supported by NSF of China, 90103002.


\begin{thebibliography}{99}
\bibitem{CDS98}A. Connes, M.R. Douglas and A. Schwarz, Noncommutative
geometry and matrix theory: compactification on tori, JHEP, {\bf 02}
(1998) 003.
\bibitem{SW99}N. Seiberg and E. Witten, String theory and noncommutative
geometry, JHEP, {\bf 09} (1999) 032.
\bibitem{Snyder47}H.S. Snyder, Quantized space-time, Phys. Rev. {\bf 71}
(1947) 38;The electromagnetic field in quantized space-time, Phys.
Rev. {\bf 72} (1947) 68.
\bibitem{Douglas01}M.M. Sheikh-Jabbari, Phys. Rev. Lett. {\bf 84} (2000)
5265;L. Bonora, M. Schnabl, M.M. Sheikh-Jabbari and A.
Tomasiello, Nucl.Phys. {\bf B589} (2000) 461;M. Chaichian, P.
Presnajder, M. M. Sheikh-Jabbari and A. Tureanu, hep-th/0107037;
A recent review to see: M.R. Douglas and N.A. Nekrasov,
Noncommutative field theory, hep-th/0106048.
\bibitem{Mw2000}S. Minwalla, M. Van Raamsdonk and N. Seiberg,
Noncommutative perturbative dynamics, JHEP {\bf 02} (2000) 020.
\bibitem{MS99}C.P. Martin and D. Sanchez-Ruiz, One-loop UV divergent
Structure of U(1) Yang-Mills Theory on Noncommutative ${\cal
R}^4$, Phys. Rev. Lett., {\bf 83} (1999) 476.
\bibitem{MM001}M.M. Sheikh-Jabbari, One Loop Renormalizability of
Supersymmetric Yang-Mills Theories on Noncommutative Two-Torus.
JHEP, {\bf 06} (1999) 015.
\bibitem{MM002}M. Chaichian, P. Presnajder, M.M. Sheikh-Jabbari and A.
Tureanu, Noncommutative Standard Model: Model Building,
hep-th/0107055.
\bibitem{MM003}M. Chaichian, A. Demichev, P. Presnajder, M.M. Sheikh-Jabbari
and A. Tureanu, Aharonov-Bohm Effect in Noncommutative Spaces,
hep-th/0012175.
\bibitem{MM004}M. Chaichian, M. M. Sheikh-Jabbari and A. Tureanu, Hydrogen Atom
Spectrum and the Lamb Shift in Noncommutative QED, Phys.Rev.Lett.
{\bf 86} (2001) 2716.
\bibitem{E821}H.N. Brown, Precise Measurement of the Positive Muon
Anomalous Magnetic Moment, Phys. Rev. Lett. {\bf 86} (2001) 2227.
\bibitem{NP}J.L. Feng and K.T. Matchev, Phys. Rev. Lett. {\bf 86} (2001) 3480;
G.-C. Cao and K. Hagiwara, Phys. Lett. {\bf B514} (2001) 123;K. Enqvist,
E. Gabrielli and K. Huitu, Phys. Lett. {\bf B512} (2001) 107;H. Baer, C. Balazs,
J. Ferrandis and X. Tata, Phys.Rev. {\bf D64} (2001) 035004.
\bibitem{SMS} V. Cirigliano, G. Ecker and H. Neufeld, Phys.Lett. {\bf B513} (2001)
361; S. Narison, Phys.Lett. {\bf B513} (2001) 53; J. Erler, and
M. Luo, Phys. Rev. Lett. 87 (2001) 071804.
\bibitem{MM005}I.F. Riad and M.M. Sheikh-Jabbari, Noncommutative QED and Anomalous
Dipole Moments, JHEP {\bf 08} (2000) 045.
\bibitem{Armoni2001}A. Armoni, Comments on perturbative dynamics of
noncommutative Yang-Mills theory, Nucl. Phys. {\bf B593} (2001)
229.
\bibitem{KT2001}V.V. Khose and G. Travaglini, Wilsonian effective action
the IR/UV mixing in noncommutative guage theories, JHEP, {\bf 01}
(2001) 026.
\bibitem{Zanon2001a}D. Zanon, Noncommutative n=1,2 super U(N) Yang-Mills:
UV/IR mixing and effective action results at one loop, Phys. Lett.
{\bf B502} (2001) 265.
\bibitem{Ruiz2001}F.R. Ruiz, Gauge-fixing independence of IR divergences
in noncommutative U(1), perturbative tachyonic instabilities and
super symmetry, Phys. Lett. {\bf B502} (2001) 274.
\end{thebibliography}
\end{document}